\documentclass[conference]{IEEEtran}
\IEEEoverridecommandlockouts
\usepackage{cite}
\usepackage{booktabs}
\usepackage{amsmath,amssymb,amsfonts}
\usepackage{algorithm}
\usepackage{graphicx}
\usepackage{textcomp}
\usepackage{xcolor}
\usepackage{algpseudocode}
\usepackage{caption}
\usepackage{braket} 
\usepackage{mdwlist}
\usepackage{caption}
\usepackage{subcaption}
\usepackage{makecell}
\usepackage{multirow}
\usepackage{bm}
\usepackage[draft]{hyperref}
\usepackage{hyperxmp}
\usepackage{balance}
\hypersetup{pdfproducer={no}}
\hypersetup{pdfcreator={no}}

\def\BibTeX{{\rm B\kern-.05em{\sc i\kern-.025em b}\kern-.08em
    T\kern-.1667em\lower.7ex\hbox{E}\kern-.125emX}}
\begin{document}

\title{Amplitude-Ensemble\\Quantum-Inspired Tabu Search Algorithm\\for Solving 0/1 Knapsack Problems\\
}

\author{
    \IEEEauthorblockN{Kuo-Chun Tseng*, Wei-Chieh Lai, I-Chia Chen, Yun-Hsiang Hsiao, Jr-Yu Chiue, and Wei-Chun Huang}
    \IEEEauthorblockA{
        \textit{Department of Computer Science and Information Engineering} \\
        \textit{National Ilan University}, Yilan, Taiwan, R.O.C. \\
        kctseng@niu.edu.tw
    }
}

\maketitle

\begin{abstract}
    In this paper, an improved version of QTS (Quantum-inspired Tabu Search) has been proposed, which enhances the utilization of population information, called ``amplitude-ensemble'' QTS (AE-QTS). This makes AE-QTS more similar to the real quantum search algorithm, Grover Search Algorithm, in abstract concept, while keeping the simplicity of the algorithm. Later, we demonstrate the AE-QTS on the classical combinatorial optimization 0/1 knapsack problem. Experimental results show that the AE-QTS outperforms other algorithms, including the QTS, by at least an average of 20\% in all cases and even by 30\% in some cases. Even as the problem complexity increases, the quality of the solutions found by our method remains superior to that of the QTS. These results prove that our method has better search performance.
\end{abstract}

\begin{IEEEkeywords}
    Quantum-inspired metaheuristic, Combinatorial optimization, Grover Search algorithm, Knapsack problem, R\&D investment
\end{IEEEkeywords}

\section{Introduction}
    Metaheuristics have been a popular research area for many years because they can be used to solve highly complex problems, including NP-complete and NP-hard problems. As a result, many researchers have devoted significant effort to this field, and have developed a number of excellent and effective metaheuristic algorithms, such as Simulated Annealing (SA) \cite{kirkpatrick1983optimization}, Genetic Algorithms (GA) \cite{holland1992adaptation}, Particle Swarm Optimization (PSO) \cite{kennedy1995particle}, Ant Colony Optimization (ACO) \cite{dorigo1996ant}, Artificial Bee Colony (ABC) \cite{karaboga2005idea}, Differential Evolution (DE) \cite{storn1997differential}, Tabu Search (TS) \cite{glover1989tabu,glover1990tabu}, and others.

    At the same time, Quantum Algorithms \cite{deutsch1991rapid,shor1994algorithms,grover1996fast} are also making remarkable achievements. The algorithms executed by quantum computers have even shaken the entire cryptographic world. This success has encouraged researchers to integrate quantum characteristics into the design of metaheuristic algorithms, giving birth to many excellent quantum-inspired algorithms such as Quantum-inspired Evolutionary Algorithm (QEA) \cite{han2002quantum}, Quantum-inspired Genetic Algorithm (QGA) \cite{narayanan1996quantum}, Quantum-inspired Ant Colony Optimization (QACO) \cite{dey2016new}, Quantum-inspired Particle Swarm Optimization (QPSO) \cite{meng2009quantum}, Quantum-inspired Differential Evolution (QDE) \cite{su2008quantum,zouache2015quantum}, Quantum-inspired Tabu Search (QTS) \cite{chou2011quantum,chiang2014quantum}, and others. These algorithms have been significantly improved by quantum characteristics.

    However, only two algorithms, QEA \cite{han2002quantum} and QTS \cite{chou2011quantum,chiang2014quantum}, are mainly conceived from the perspective of quantum properties. The idea of QEA \cite{han2002quantum} is still slightly more inclined to the traditional population-based thinking mode, but the concept of improving the observation probability in quantum algorithms is perfectly mirrored in the thinking of quantum bit measurement and updating quantum state. This feature is even more fully utilized in the QTS algorithm \cite{chou2011quantum,chiang2014quantum}. QTS \cite{chou2011quantum,chiang2014quantum} perfectly mirrors the concept of the entire Grover Search Algorithm \cite{grover1996fast}. By comparing each bit of the best and worst solutions in each iteration of the population, the trend of each bit can be known. This trend adjusts the quantum bit's state via a rotation matrix, similar to solution adjustment probabilities in Grover Search Algorithm \cite{grover1996fast}.

    Quantum-inspired algorithms that are derived from pure quantum algorithms \cite{deutsch1991rapid,shor1994algorithms,grover1996fast} are relatively simple and unambiguous in their structure and implementation. Unlike other metaheuristic methods, which have many different implementation methods for various components, it is difficult to reproduce experiments or apply them to other problems. Among quantum-inspired algorithms, QTS \cite{chou2011quantum,chiang2014quantum} is the simplest and easiest to implement. All components of its algorithm have standardized implementation methods, with no ambiguity. Experimental results also show that QTS \cite{chou2011quantum,chiang2014quantum} is not only simple to implement and fast in computation, but also superior to QEA \cite{han2002quantum} in terms of performance. It can find the optimal solution more quickly and stably.

    Although the idea of QTS \cite{chou2011quantum,chiang2014quantum} is already superior, its performance has not yet been maximized. The solutions generated by each iteration of the population only screen out two sets of solutions to update the quantum state. Obviously, other sets of solutions are wasted, but these solutions are also generated by quantum bits containing all the information. This information should also be incorporated into all quantum bits. Similar to the Grover Search Algorithm \cite{grover1996fast}, if the amplitude of a solution is to be adjusted, the amplitude of all other sets of solutions must be considered. Therefore, it is very important to find a way to put the information of the remaining population back into the quantum bits.

\section{Controbutions}
    This section elaborates in detail the contributions of this study. In fact, the QTS algorithm \cite{chou2011quantum,chiang2014quantum} is already highly efficient and simplistic, making it quite challenging to enhance the essence of the algorithm further. At most, improvements can be made by adding external components when implementing it for different problems to cater to their specific needs. Remarkably, our study manages to enhance the nature of the QTS algorithm \cite{chou2011quantum,chiang2014quantum} and achieves exceptional efficiency, a feat that is not easily accomplished. The contributions of our research are summarized as follows:
    \begin{enumerate}
        \item Our AE-QTS, modified from the algorithm design layer, achieves at least an average 20\% performance improvement and up to 30\% in some cases. The quality of the solutions found is also superior to QTS when facing more complex problems. Additionally, this advancement is compatible with issues previously addressed using QTS \cite{chou2011quantum,chiang2014quantum}; requiring only a modification to the \textbf{update} quantum state part.
        \item Maintain the simplicity and ease of implementation of the QTS algorithm \cite{chou2011quantum,chiang2014quantum}. QTS \cite{chou2011quantum,chiang2014quantum} is different from other traditional metaheuristics in that it has easy implementation characteristics. This is because each component of the algorithm has no ambiguity, and there are no different interpretations or programming methods in the implementation. Consequently, it doesn't spawn different adaptations or extensions, ensuring a rapid and straightforward implementation that exhibits consistent search capability across diverse problems.
        \item AE-QTS retains the original number of parameters of QTS \cite{chou2011quantum,chiang2014quantum}. Often, many improvement methods introduce additional parameters that require adjustment. However, this particular method of improvement does not necessitate any extra parameters. Consequently, re-testing the parameters or adjusting other variables is unnecessary.
    \end{enumerate}

\section{Quantum Computing Principles}
    The basic unit of a quantum computer is the quantum bit or qubit. Qubits are wave-like, so each state has a corresponding amplitude. Therefore, all states have their own probability of being observed. When a qubit is in this state and is used for computation, it can carry infinite information. This property can be used to achieve the goal of parallel computing. However, the most challenging part is not designing the calculation method but designing the adjustment of its amplitude to increase the probability of the target state being observed. The two quantum-inspired algorithms currently available are both designed based on this concept. Next, we will introduce qubits and their corresponding logic gates.
   
    \subsection{Quantum bit}
    A quantum bit, or qubit, is a unit of quantum information. It can be in one of two states, which are represented as $\ket{0}$=$\begin{pmatrix} 1 \\ 0 \end{pmatrix}$ and $\ket{1}$=$\begin{pmatrix} 0 \\ 1 \end{pmatrix}$. These states can also be viewed as the $\hat{x}$ and $\hat{y}$ axes of a two-dimensional plane. Unlike classical bits in either the 0 or 1 state, a qubit can exist in a superposition of both $\ket{0}$ and $\ket{1}$ states simultaneously, referred to as a superposition state. The superposition state can be seen as a point on the circle in the two-dimensional plane, as shown in Fig.~\ref{fig:polar}. Therefore, it can be expressed as:
    $\ket{\phi}=\alpha\ket{0}+\beta\ket{1}$,
    where 
    $\left| \alpha \right|^{2}$ represents the probability of obtaining $\ket{0}$ when measuring this qubit, $\left| \beta \right|^{2}$ represents the probability of obtaining $\ket{1}$ when measuring this qubit, and $\left| \alpha \right|^{2}+\left| \beta \right|^{2}=1$.
        
    \subsection{Quantum gate}
\begin{figure}[h]
            \centerline{\includegraphics[scale=0.4]{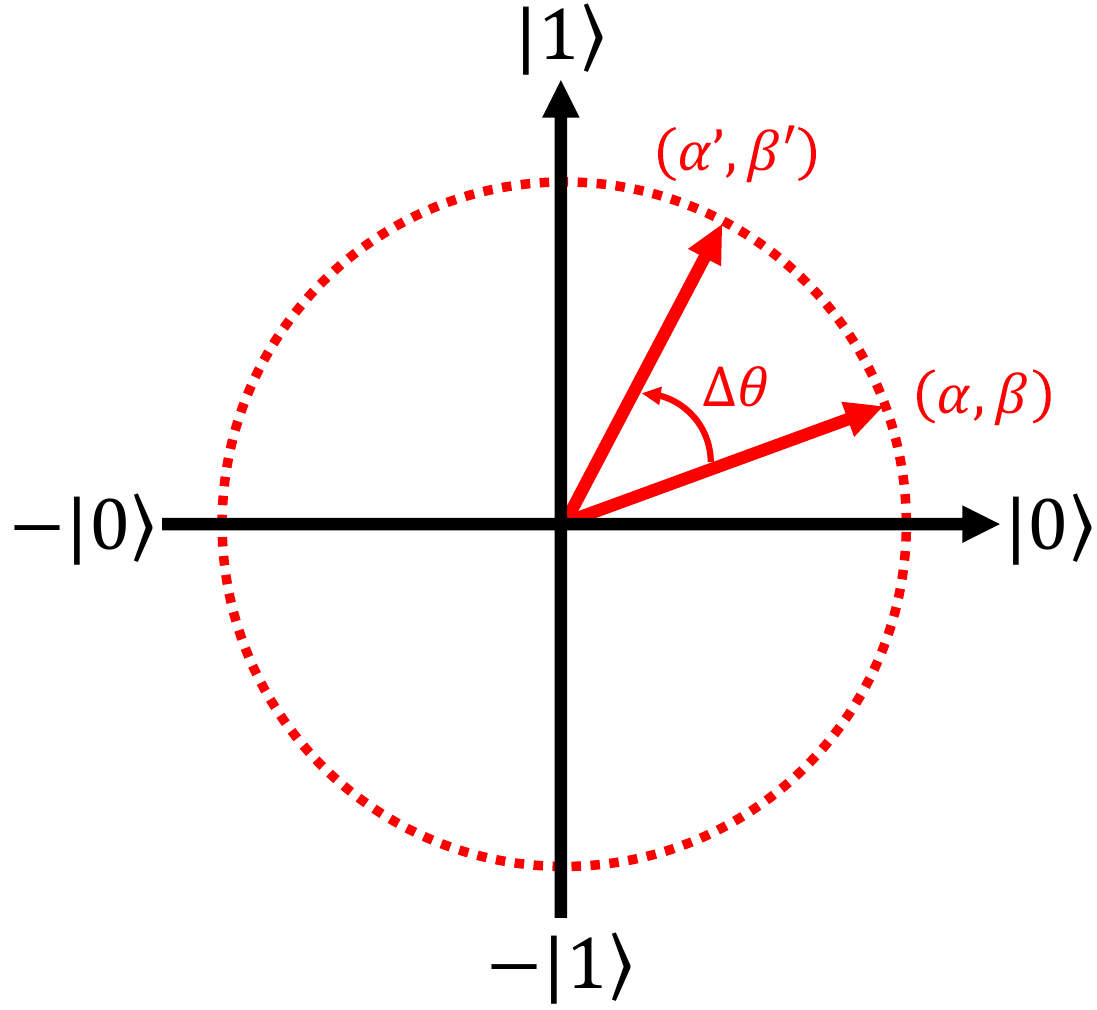}}
            \caption{Polar plot of the rotation matrix on a qubit.}
            \label{fig:polar}
        \end{figure}

        The distinction between quantum logic gates and classical computer logic gates lies in the reversibility of the former, as they all perform unitary operations, denoted as $UU^{*}=U^{*}U=I$. There are numerous types of quantum logic gates; however, this study exclusively utilizes rotation matrices.
        \begin{equation}
            \binom{\alpha^{'}}{\beta^{'}}=
            \begin{pmatrix}
                \cos(\bigtriangleup \theta) & -\sin(\bigtriangleup \theta) \\
                \sin(\bigtriangleup \theta) & \cos(\bigtriangleup \theta) 
            \end{pmatrix}
            \binom{\alpha}{\beta}
        \end{equation}
        Referring to the Fig.~\ref{fig:polar}, $\alpha^{'}$ and $\beta^{'}$ represent the coordinates after the rotation, and the probabilities of obtaining $\ket{0}$ and $\ket{1}$ upon measurement of the quantum bit can be modified through the rotation matrix. It is essential to note that the operation involves clockwise rotation. Consequently, in the second and fourth quadrants, rotation is executed using $-\bigtriangleup\theta$.
        
\section{0/1 knapsack problem}
    The 0/1 knapsack problem is a classic combinatorial optimization problem. Generally, there are $K$ items to choose from, each of which can either be taken or left behind. Each item has a specific profit and weight. The knapsack also has a weight limit $C$, making it impossible to take all items. This problem aims to maximize the total profit of the items taken while adhering to the weight constraint of the knapsack. Here, we define the knapsack problem as adopted in this study: $p_i=$ profit of item $i$, $w_i=$ weight of item $i$, and $C=\frac{1}{2}\sum_{i=1}^{K}w_i$.\\
    The objective of this problem is to maximize
    \begin{equation*}  
        f(x)=\sum_{i=1}^{K}p_ix_i,
    \end{equation*}
    subject to the constraint
        \begin{equation*}
            \sum_{i=1}^{K}w_ix_i \le C.
        \end{equation*}
    Here, $x=\left(x_1, x_2, ..., x_k\right)$, with $x_i$ being 0 or 1, $p_i$ representing the profit of the $i$th item, and $w_i$ denoting the weight of the $i$th item. $C$ is the maximum capacity of the knapsack. If $x_i=1$, it indicates that the $i$th item has been selected to be placed in the knapsack.

    This problem can be simply categorized into three cases, each with different methods of generating weights and profits, as follows:
    \begin{basedescript}{
        \desclabelstyle{\multilinelabel}
        \desclabelwidth{1.4cm}}
        \item[Case I:] The weight $w_i$ is uniformly random selected from the range $\left[1, 10\right]$, and the profit $p_i$ is calculated as $w_i+5$.
        \item[Case II:] The weight $w_i$ is uniformly random selected from the range $\left[1, 10\right]$, and the profit $p_i$ is calculated as $w_i+l_i$, where $l_i$ is uniformly random selected from the range $\left[0, 5\right]$.
        \item[Case III:] The weight $w_i$ is within the range $\left\{ x \in \mathbb{Z} : 1\le x \le 10 \right\}$. The profits of $w_i$ are generated sequentially and in a cyclic manner, for example, $w=\left(1, 2, ..., 10, 1, 2, ...\right)$. The profit $p_i$ is calculated as $w_i+5$.
    \end{basedescript}
    
\section{Amplitude-ensemble QTS (AE-QTS)}
    This method is derived from QTS \cite{chou2011quantum,chiang2014quantum}, where QTS selects the best and worst from each iteration to determine the angle rotation of each qubit. It's a highly effective strategy, and experiments have confirmed its capability to adjust the amplitude of quantum bits precisely. In this study, we adopt this method but expand its scope to encompass the entire population of each iteration. Doing so enables a more efficient incorporation of all explored information into the qubits.

    Thus, \textbf{the only difference between this study and QTS \cite{chou2011quantum,chiang2014quantum} lies in the approach at line 9 of Algo.~\ref{alg:AE-QTS}}; all other steps are the same. The explanations are as follows.

\begin{figure}[h]
        \centerline{\includegraphics[scale=0.5]{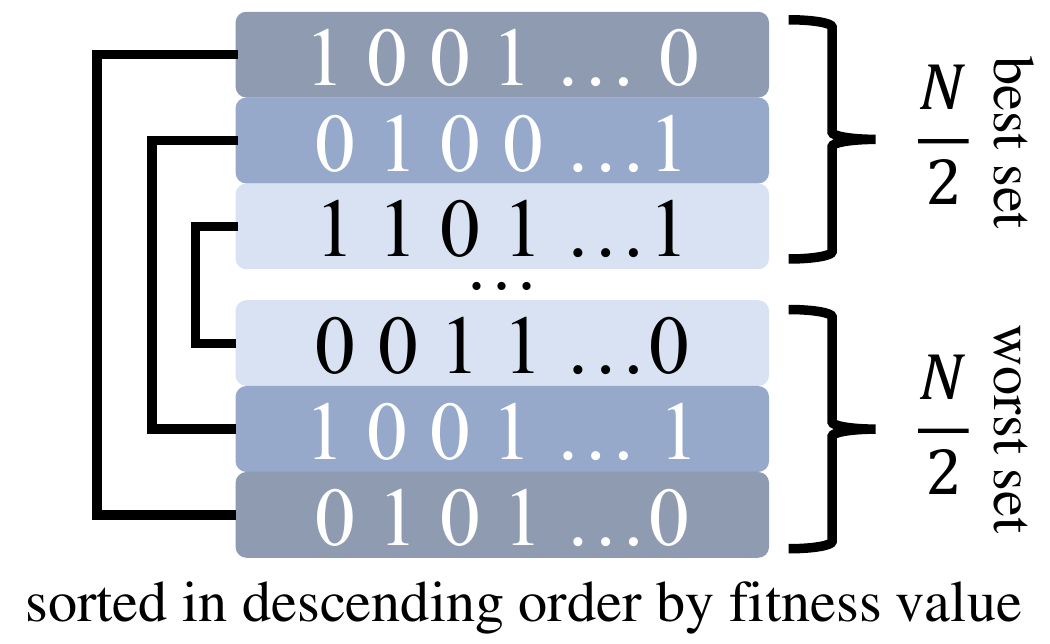}}
        \caption{The selection pair in $P(t)$.}
        \label{fig:update_pair}
    \end{figure}

    \begin{enumerate}
        \item Set the iteration $t$ to be 0.
        \item $Q(0)=\left(
            \begin{array}{c|c|c|c}
                \begin{matrix}\alpha_1 \\ \beta_1 \end{matrix} &
                \begin{matrix}\alpha_2 \\ \beta_2 \end{matrix} &
                \begin{matrix}... \\ ... \end{matrix} &
                \begin{matrix}\alpha_k \\ \beta_k \end{matrix}
            \end{array}
            \right)$, where $k$ is the number of qubits, equivalent to the number of items with the knapsack problem. During the initialization phase, all values of $\alpha$ and $\beta$ are set to $\frac{1}{\sqrt{2}}$. This implies that when measuring these quantum bits, there is an equal probability of measuring $\ket{0}$ or $\ket{1}$.
        \item The best solution $s^b$ is selected from $P(t)$, which was measured from $Q(t)$, repaired, and evaluated again within $P(t)$.
        \item The portion from the $4th$ line to the $14th$ line constitutes the core of this method, which will be executed repeatedly until reaching the maximum number of iterations.
        \item \textbf{make} $P(t)$ implies generating $N$ sets of solutions by measuring $Q(t-1)$ for $N$ times, where these solutions are represented in binary encoding, signifying the selection or non-selection of items. For example, if $N=2$, $K=3$, and $Q(t-1)=\left( \begin{array}{c|c|c}
            \begin{matrix}\alpha_1 \\ \beta_1 \end{matrix} &
            \begin{matrix}\alpha_2 \\ \beta_2 \end{matrix} &
            \begin{matrix}\alpha_3 \\ \beta_3 \end{matrix}
            \end{array} \right)$. In this case, 6 measurements are needed to generate 2 sets of binary solutions with a length of 3. The probabilities of obtaining `0' and `1' are $\left| \alpha_k \right|^2$ and $\left| \beta_k \right|^2$, where $k=1, 2, 3$, respectively, as shown below:
            $\begin{pmatrix}
            1 & 0 & 0 \\
            1 & 1 & 0
            \end{pmatrix}$
        \item \textbf{repair} $P(t)$ (Algo.~\ref{alg:repair}) means that when the total weight exceeds the capacity of the knapsack, it is necessary to perform repairs by removing some items to reduce the weight. However, if the total weight is less than the knapsack's capacity, attempts can be made to add some items back, while trying to maximize the knapsack's carrying capacity during the repair process as much as possible.
        \item \textbf{evaluate} $P(t)$ means assessing the profits obtained for each set of selections. There are a total of $N$ feasible solutions here, so it will return the profits corresponding to the $N$ feasible solutions.
        \item \textbf{update} $Q(t-1)$ (Algo.~\ref{alg:update}) means updating the state of quantum bits. This is the only difference from QTS. In QTS, the contemporary best and worst solutions are taken from $P(t)$, and a comparison is made using the corresponding bits. The rotation angles for each bit are determined according to Table~\ref{tbl:rotate_table} for the rotation matrix operation. In this study, the population is sorted based on profit. Then, pairs are selected as follows: the best and worst, the second best and second worst, the third best and third worst, and so on, up to $N/2$ pairs (shown in Fig.~\ref{fig:update_pair}). The first pair is operated on with $\Delta\theta/1$ for the rotation matrix, the second pair with $\Delta\theta/2$, the third pair with $\Delta\theta/3$, and so on, following this pattern. Quantum state adjustments are made using these pairs, according to Table~\ref{tbl:rotate_table}. In total, $N/2$ rotation matrix operations are performed.
        \item The next step is to evaluate whether the best solution found in this iteration of the population is better than the currently recorded best solution. If it is, a replacement is made. Consequently, this measure ensures that $s^b$ always represents the best binary solution found throughout the algorithm's execution.
    \end{enumerate}
    
    \begin{algorithm}
        \caption{\textbf{Amplitude-Ensemble QTS}}\label{alg:AE-QTS}
        \begin{algorithmic}[1]
            \State $t \gets 0$
            \State initialize $Q(t)$
            \State initialize the best solution $s^b$ from $P(t)$ by measuring $Q(t)$, repairing and evaluating $P(t)$
            \While{$t <$ MAX\_ITER}
                \State $t \gets t+1$
                \State \textbf{make} $P(t)$ with $N$ solutions by measuring $Q(t-1)$
                \State \textbf{repair} $P(t)$
                \State \textbf{evaluate} $P(t)$ 
                \State \textbf{update} $Q(t)$ by $P(t)$ \textcolor{red}{\Comment{The difference from QTS}}
                \State select the best solution $b$ from $P(t)$
                \If{$b > s^{b}$}
                    \State $s^{b} \gets b$
                \EndIf
            \EndWhile
        \end{algorithmic}
    \end{algorithm}

    \begin{algorithm}
        \caption{\textbf{Repair($s$)}}\label{alg:repair}
        \begin{algorithmic}[1]
            \State $s=\left( x_1, x_2, ..., x_k \right)$
            \State knapsack-overfilled $\gets$ false
            \If{$\sum_{i=1}^{K} w_ix_i > C$}
                \State knapsack-overfilled $\gets$ true
            \EndIf
            \While{knapsack-overfilled}
                \State random select $j$th item from $s$
                \State $s_j \gets 0$
                \If{$\sum_{i=1}^{K} w_ix_i \le C$}
                    \State knapsack-overfilled $\gets$ false
                \EndIf
            \EndWhile
            \While{not knapsack-overfilled}
                \State random select $j$th item from $s$
                \State $s_j \gets 1$
                \If{$\sum_{i=1}^{K} w_ix_i > C$}
                    \State knapsack-overfilled $\gets$ true
                \EndIf
            \EndWhile
        \end{algorithmic}
    \end{algorithm}

    \begin{algorithm}
        \caption{\textbf{Update($q$)}}\label{alg:update}
        \begin{algorithmic}[1]
            \State $i \gets 0$
            \State $index \gets$ \textbf{sort} $P(t)$ with profits by DESC
            \While{$i < N/2$}
                \State $best \gets index[i]$
                \State $worst \gets index[N-1-i]$
                \State $j \gets 0$
                \While{$j < K$}
                    \State $t \gets i+1$
                    \State Determine $\Delta\theta$ by $best_j$ and $worst_j$ in Table~\ref{tbl:rotate_table}
                    \State $q_j=\begin{pmatrix}
                        cos(\Delta\theta/t) & -sin(\Delta\theta/t) \\
                        sin(\Delta\theta/t) & cos(\Delta\theta/t)
                    \end{pmatrix}
                    \begin{pmatrix}
                    \alpha_j \\
                    \beta_j
                    \end{pmatrix}$
                    \State $j \gets j+1$
                \EndWhile
                \State $i \gets i+1$
            \EndWhile
        \end{algorithmic}
    \end{algorithm}

    \begin{table}[ht]
    \caption{Rotation lookup table}\label{tbl:rotate_table}
    \centering
        \begin{tabular}{llll}
            \hline
            $s^b_i$ & $s^w_i$ & $Q_i \in T$ & $\Delta\theta$ \\ \hline
            \multicolumn{4}{l}{$Q_i$ locates in first or third quadrant} \\
              0  &  0   &  True   &  0  \\
              0  &  1   &  False  &  $-\theta$  \\
              1  &  0   &  False  &  $+\theta$  \\
              1  &  1   &  True   &  0  \\
            \multicolumn{4}{l}{$Q_i$ locates in second or fourth quadrant} \\
              0  &  0   &  True   &  0  \\
              0  &  1   &  False  &  $+\theta$  \\
              1  &  0   &  False  &  $-\theta$  \\
              1  &  1   &  True   &  0  \\ \hline
            \multicolumn{4}{l}{$i=1, 2,..., k,$ where $k$ is the number of qubit.}\\
            \multicolumn{4}{l}{$T = \{x\ |\ \neg(s^b_i \oplus s^w_i)\}$, where 1 (0) means the qubit is (not) tabooed.}
        \end{tabular}
    \end{table}

\section{Experiments}
    Here, we use the abbreviation AE-QTS to refer to our method. The experiments were implemented using Python, and we compared our method with several classical algorithms, including GA, TS, DE, ABC, QEA \cite{han2002quantum}, and QTS \cite{chou2011quantum,chiang2014quantum}. To ensure fairness, \textbf{we limited the population size ($\bm{population\_size = 10}$ or $\bm{N=10}$) and maximum number of iterations ($\bm{MAX\_ITER=1000}$) for all algorithms to ensure that all algorithms have the same number of evaluations ($\bm{population\_size*MAX\_ITER=10000}$).} We also ran all algorithms with optimal parameters, including GA, TS, DE, ABC, and $\Delta\theta=0.01\pi$ for QEA \cite{han2002quantum}, QTS \cite{chou2011quantum,chiang2014quantum}, and AE-QTS.

\begin{figure*}
        \centering
        \begin{subfigure}[b]{0.3\textwidth}
            \centering
            \includegraphics[width=\textwidth]{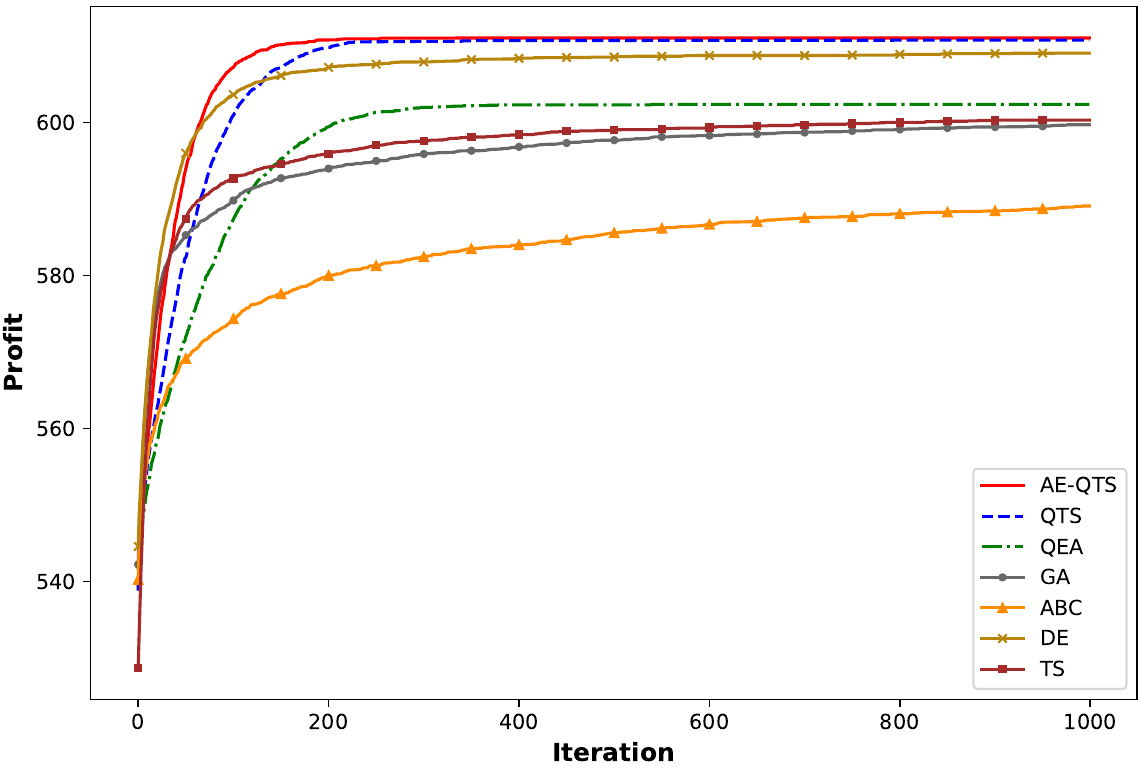}
            \caption{100 items}
            \label{fig:casei_100}
        \end{subfigure}
        \hfill
        \begin{subfigure}[b]{0.3\textwidth}
            \centering
            \includegraphics[width=\textwidth]{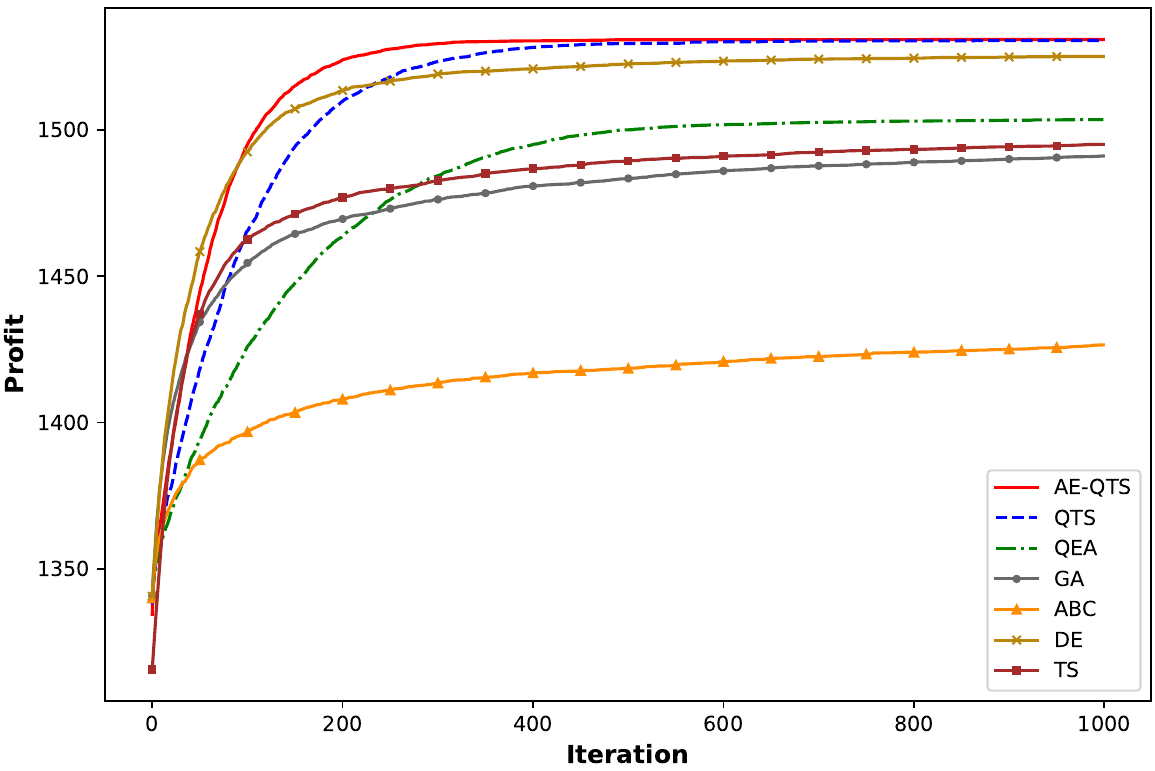}
            \caption{250 items}
            \label{fig:casei_250}
        \end{subfigure}
        \hfill
        \begin{subfigure}[b]{0.3\textwidth}
            \centering
            \includegraphics[width=\textwidth]{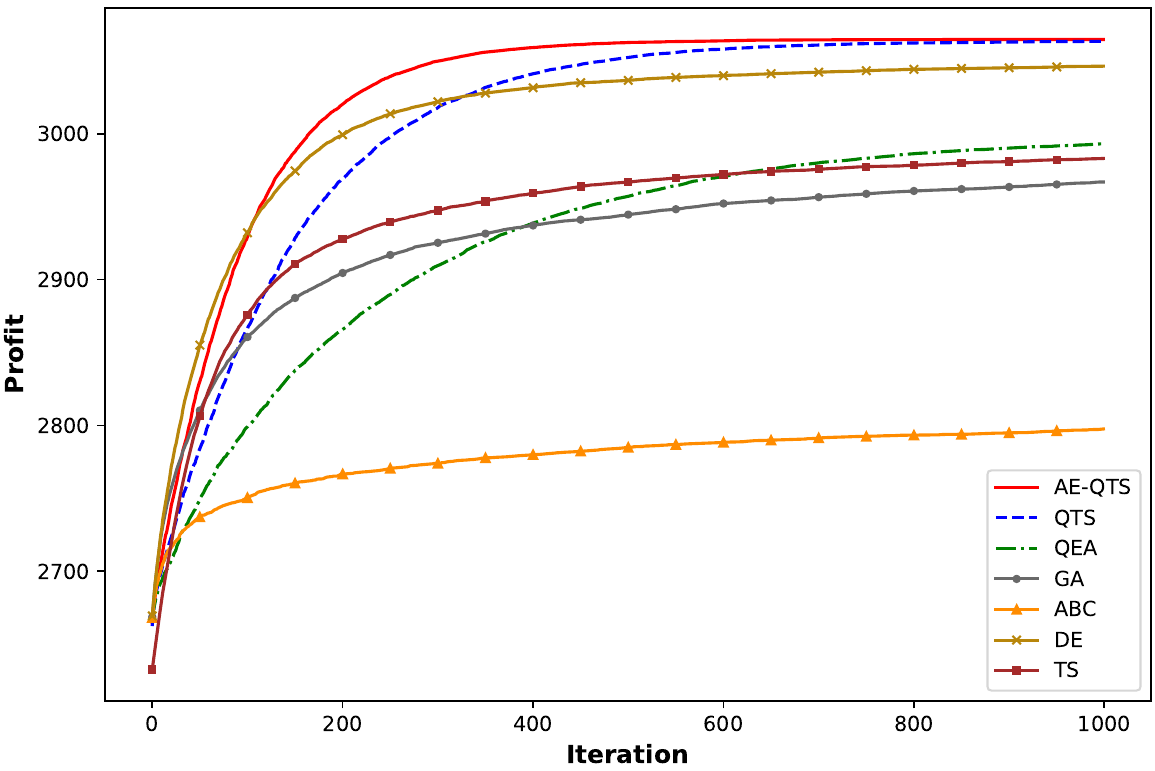}
            \caption{500 items}
            \label{fig:casei_500}
        \end{subfigure}
        \caption{Case I: 100, 250, and 500 items.}
        \label{fig:casei}
    \end{figure*}

\begin{figure*}
        \centering
        \begin{subfigure}[b]{0.3\textwidth}
            \centering
            \includegraphics[width=\textwidth]{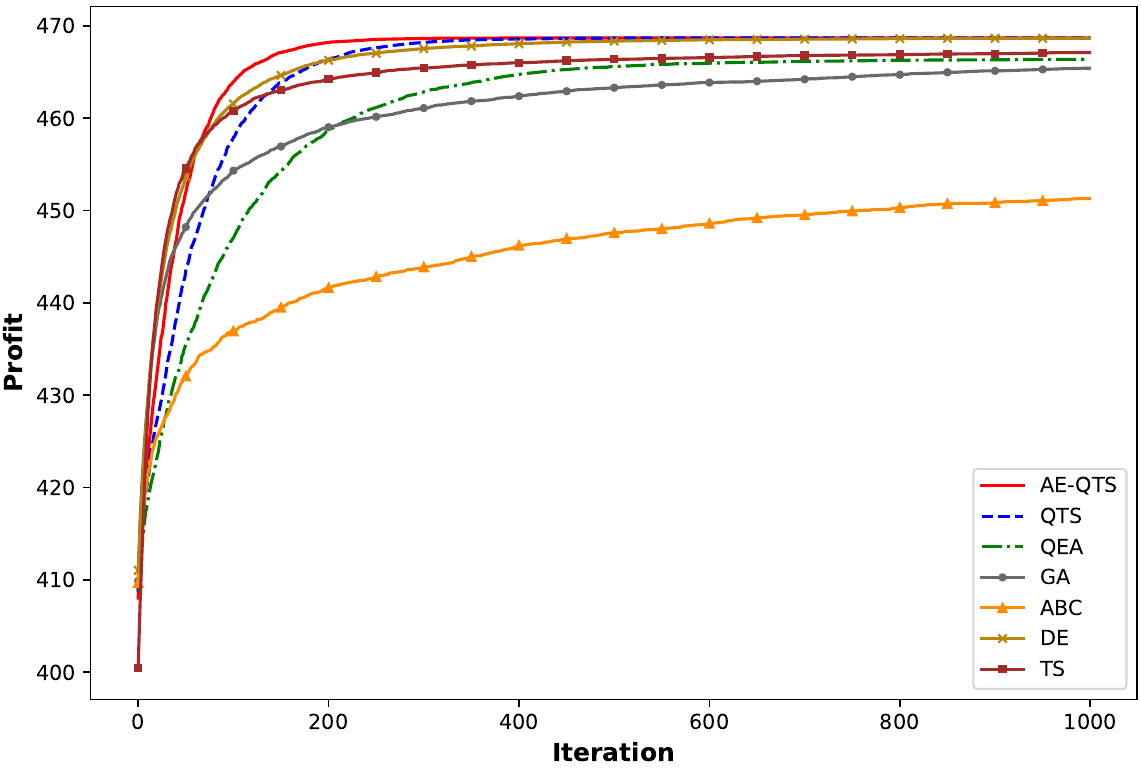}
            \caption{100 items}
            \label{fig:caseii_100}
        \end{subfigure}
        \hfill
        \begin{subfigure}[b]{0.3\textwidth}
            \centering
            \includegraphics[width=\textwidth]{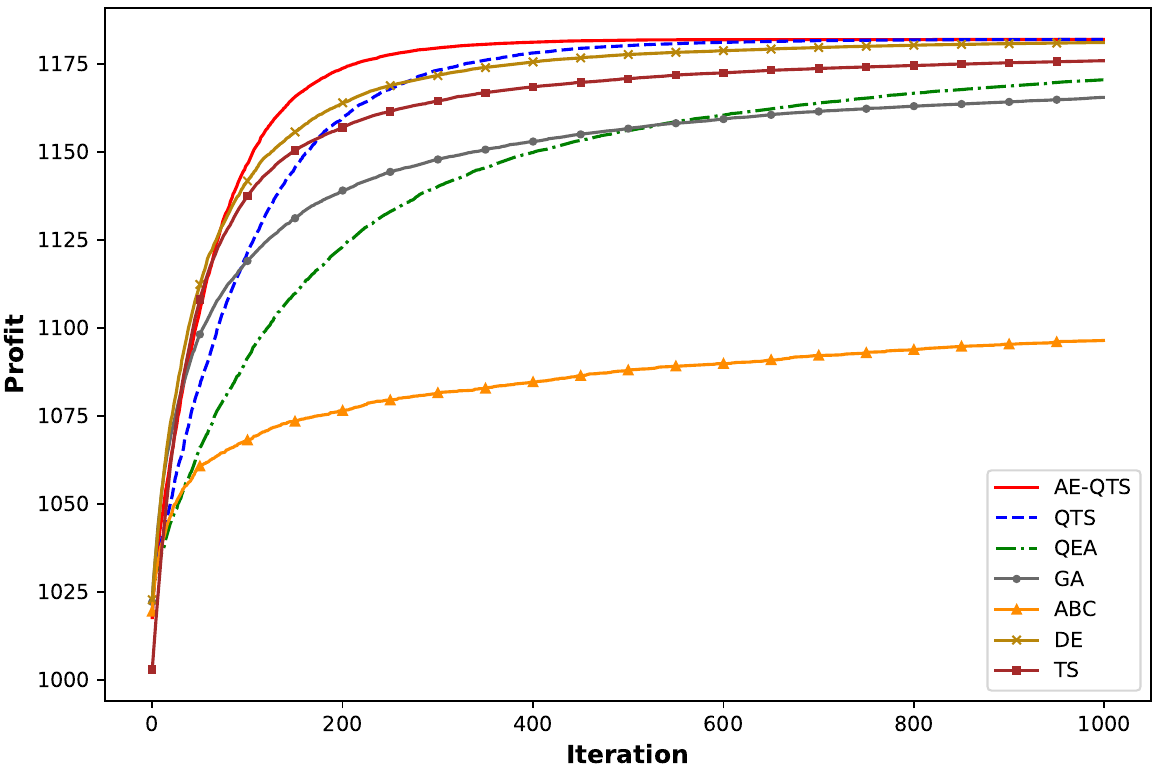}
            \caption{250 items}
            \label{fig:caseii_250}
        \end{subfigure}
        \hfill
        \begin{subfigure}[b]{0.3\textwidth}
            \centering
            \includegraphics[width=\textwidth]{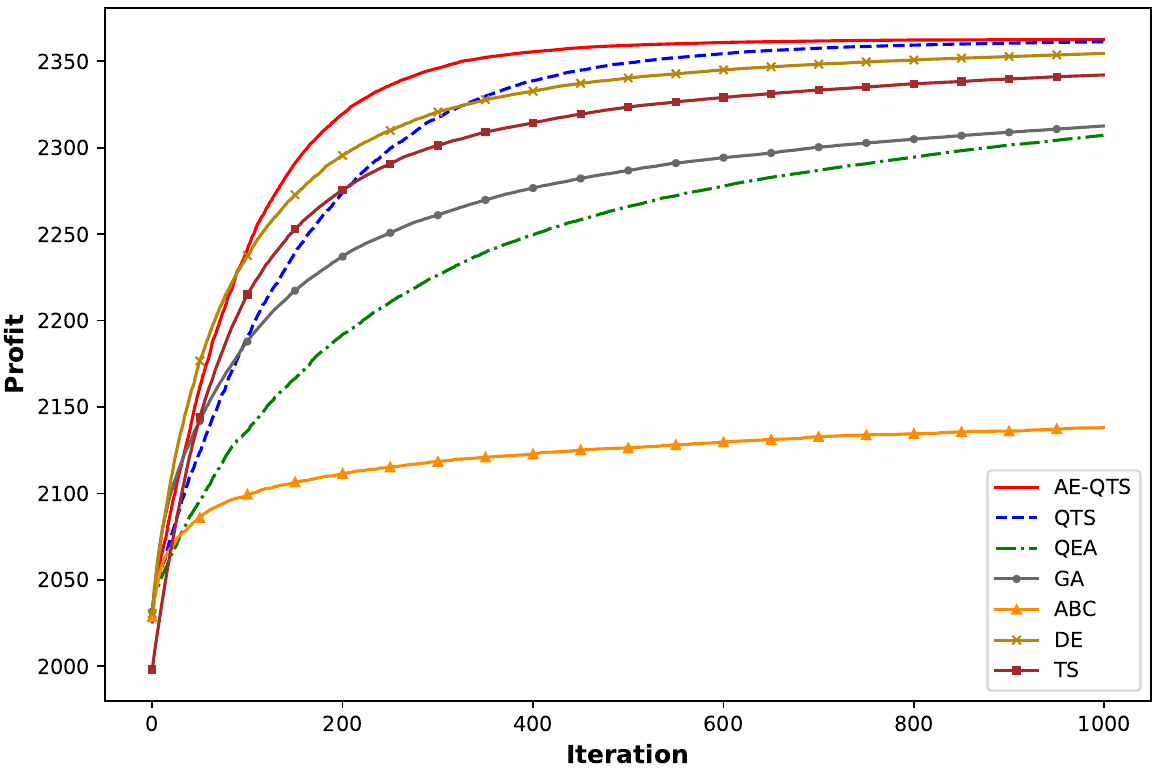}
            \caption{500 items}
            \label{fig:caseii_500}
        \end{subfigure}
        \caption{Case II: 100, 250, and 500 items.}
        \label{fig:caseii}
    \end{figure*}

\begin{figure*}
        \centering
        \begin{subfigure}[b]{0.3\textwidth}
            \centering
            \includegraphics[width=\textwidth]{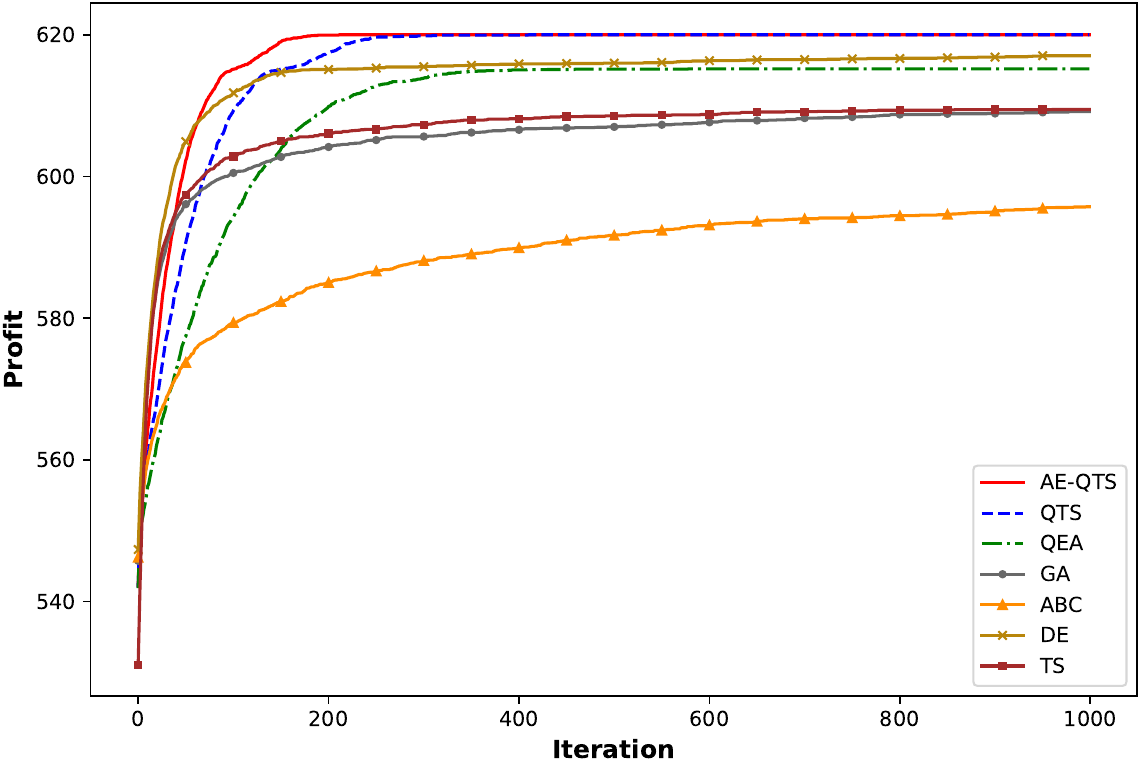}
            \caption{100 items}
            \label{fig:caseiiI_100}
        \end{subfigure}
        \hfill
        \begin{subfigure}[b]{0.3\textwidth}
            \centering
            \includegraphics[width=\textwidth]{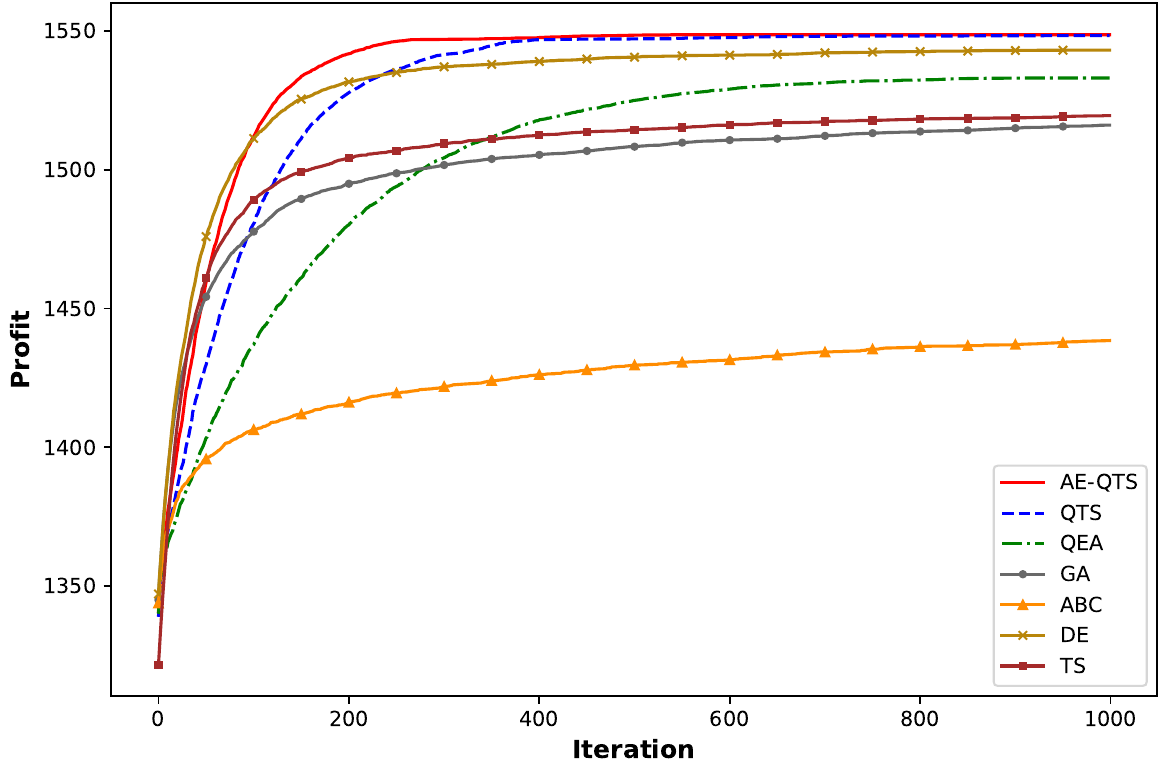}
            \caption{250 items}
            \label{fig:caseiii_250}
        \end{subfigure}
        \hfill
        \begin{subfigure}[b]{0.3\textwidth}
            \centering
            \includegraphics[width=\textwidth]{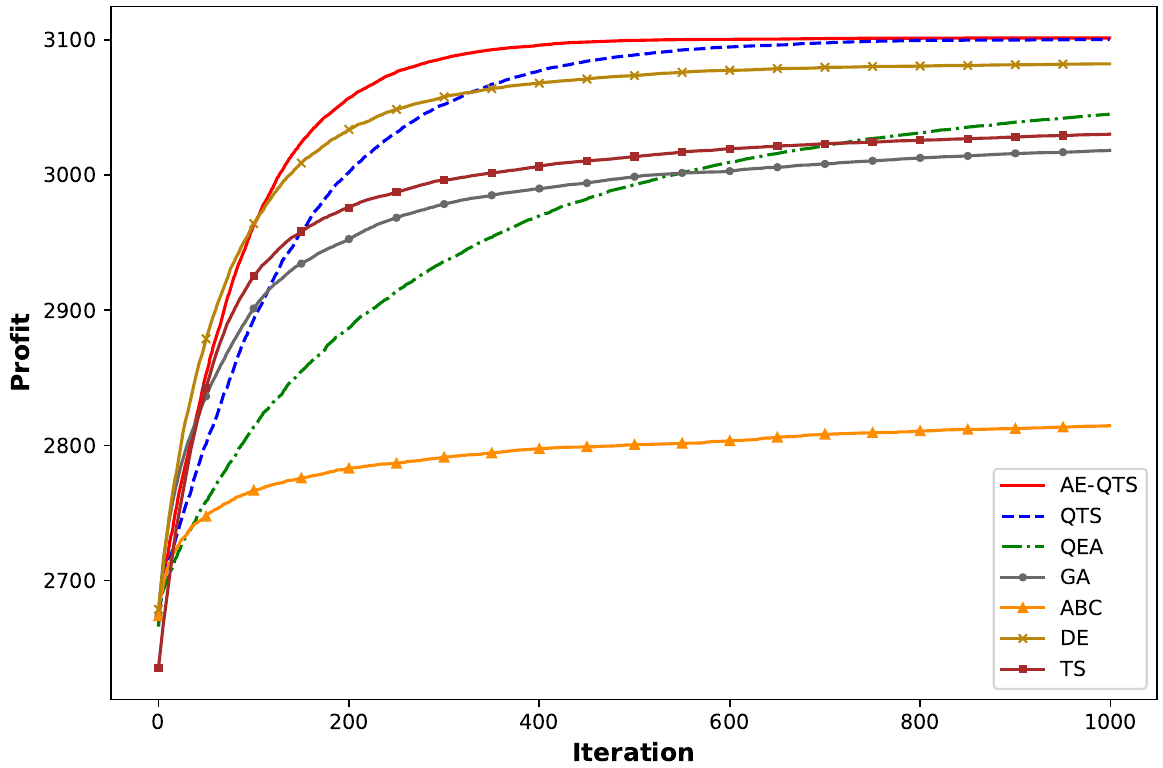}
            \caption{500 items}
            \label{fig:caseiii_500}
        \end{subfigure}
        \caption{Case III: 100, 250, and 500 items.}
        \label{fig:caseiii}
    \end{figure*}

    The subsequent experiments were all run 100 times and averaged. The experiments were conducted with 100, 250, and 500 items. The experimental results are shown in Fig.~\ref{fig:casei}, Fig.~\ref{fig:caseii}, and Fig.~\ref{fig:caseiii}. These nine experimental results show that QTS \cite{chou2011quantum,chiang2014quantum} and AE-QTS have the best convergence results in the end. QTS \cite{chou2011quantum,chiang2014quantum} initially has a relatively slow convergence speed, but the final convergence results are excellent. AE-QTS has a convergence speed close to DE at the beginning, and the convergence results are as good as QTS \cite{chou2011quantum,chiang2014quantum} in the later stage. This means that AE-QTS is better than QTS \cite{chou2011quantum,chiang2014quantum} and is even the best algorithm in the comparison. 

\begin{figure*}
        \centering
        \begin{subfigure}[b]{0.3\textwidth}
            \centering
            \includegraphics[width=\textwidth]{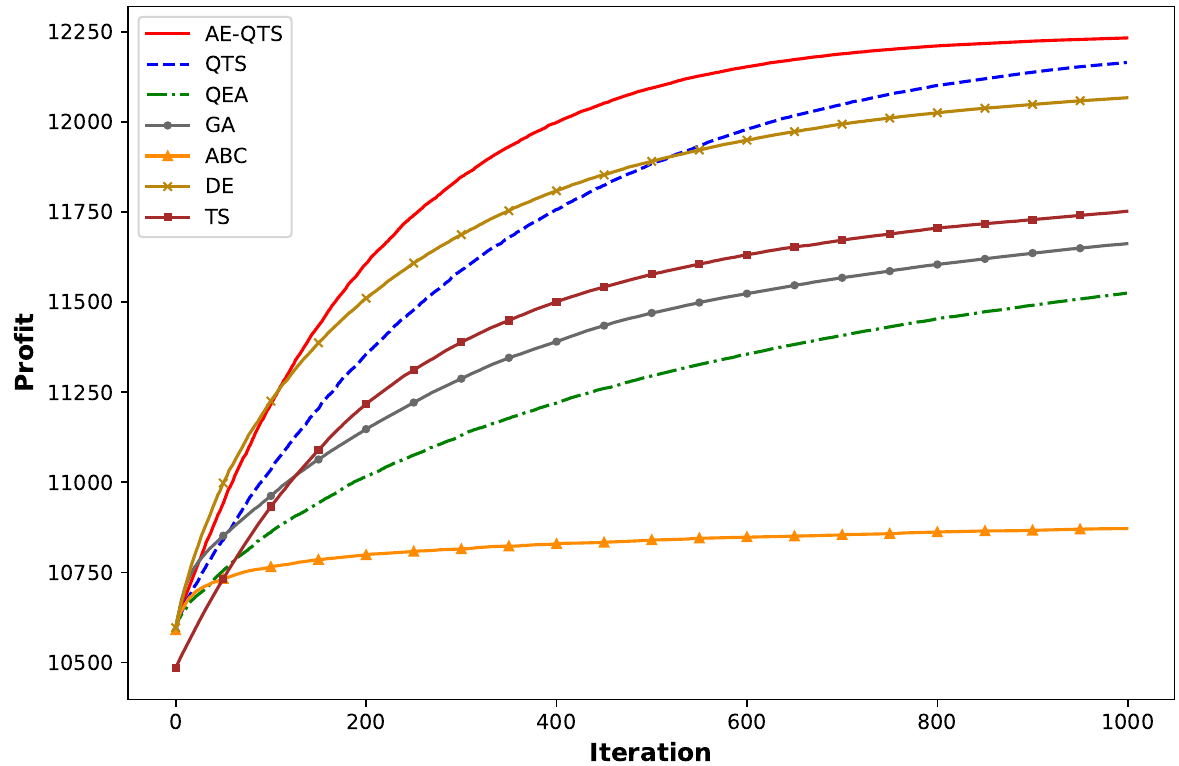}
            \caption{Case I}
            \label{fig:casei2000item}
        \end{subfigure}
        \hfill
        \begin{subfigure}[b]{0.3\textwidth}
            \centering
            \includegraphics[width=\textwidth]{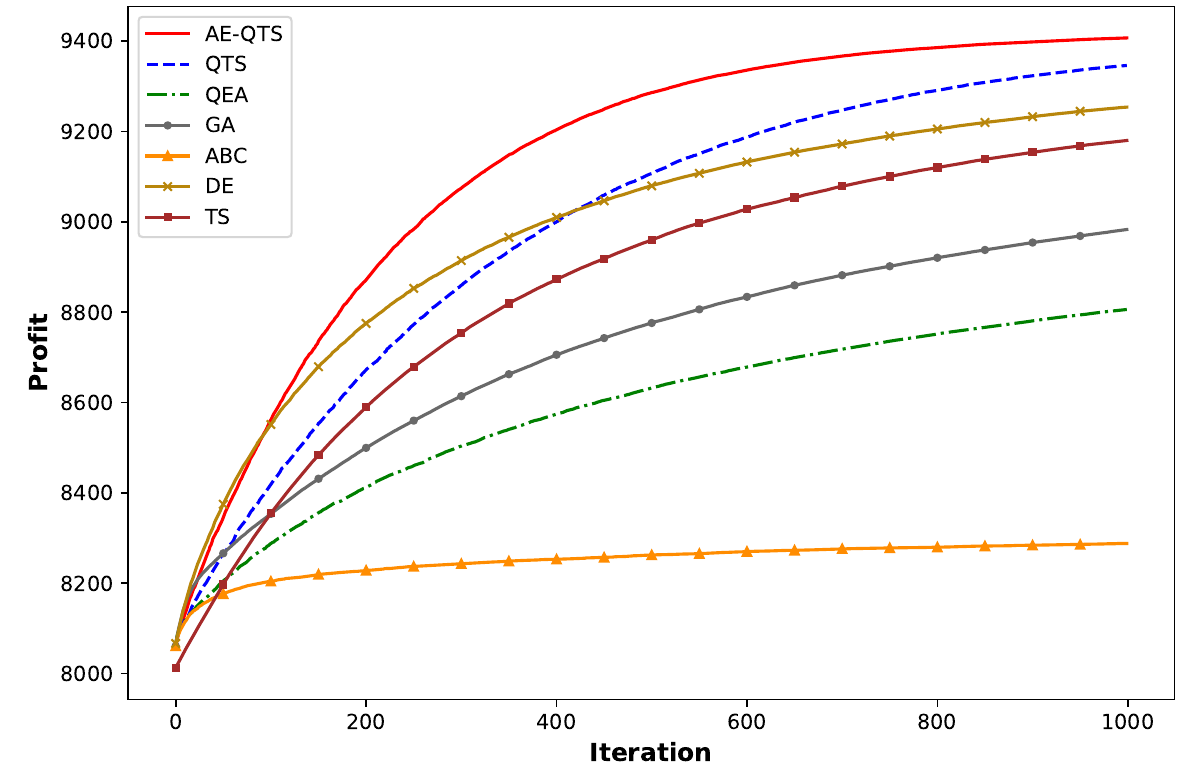}
            \caption{Case II}
            \label{fig:caseii2000item}
        \end{subfigure}
        \hfill
        \begin{subfigure}[b]{0.3\textwidth}
            \centering
            \includegraphics[width=\textwidth]{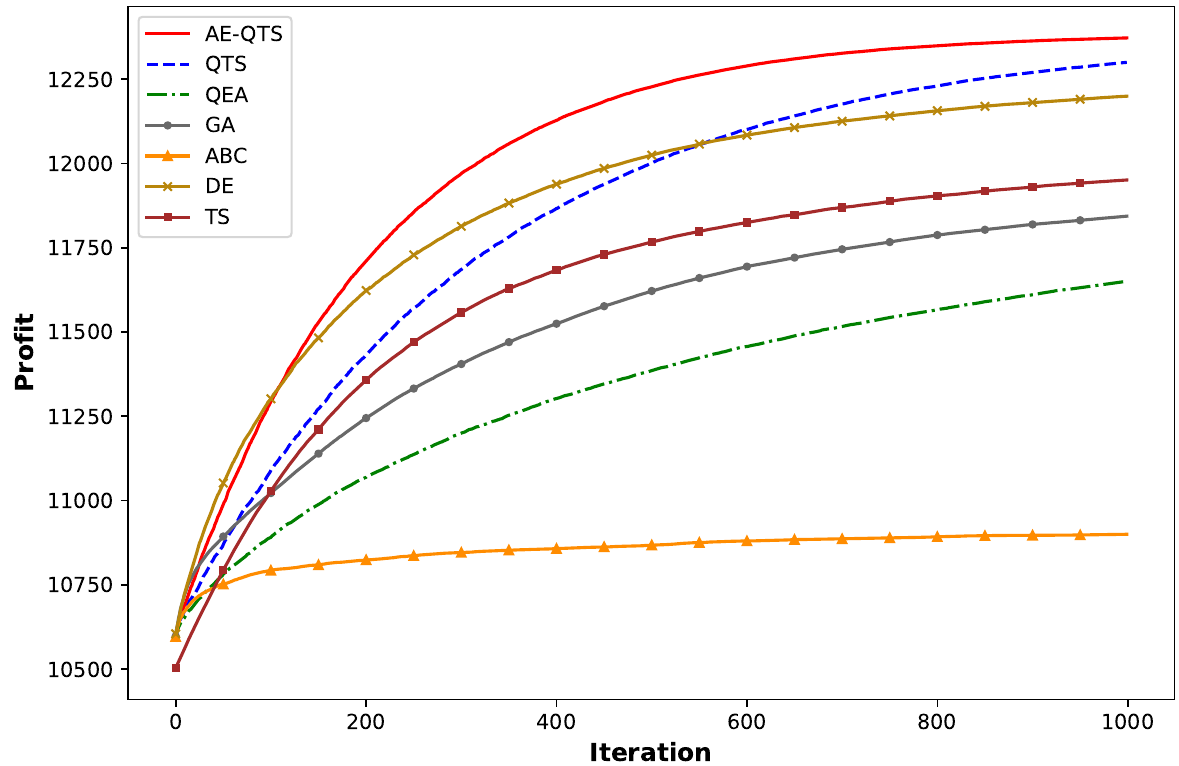}
            \caption{Case III}
            \label{fig:caseiii2000item}
        \end{subfigure}
        \caption{Three cases with 2000 items.}
        \label{fig:allcases}
    \end{figure*}

    When there are 500 items, QTS \cite{chou2011quantum,chiang2014quantum} can converge to a solution that is as good as AE-QTS. Therefore, we increase the problem difficulty for testing. We use 2000 items for testing, and the results are shown in Fig.~\ref{fig:allcases}. It can be found that AE-QTS has the best convergence results among all algorithms. It even converges faster than other algorithms. Its convergence results are significantly better than other algorithms, and the results it finds are also better than other algorithms.

\begin{figure*}
        \centering
        \begin{subfigure}[b]{0.45\textwidth}
            \centering
            \includegraphics[width=\textwidth]{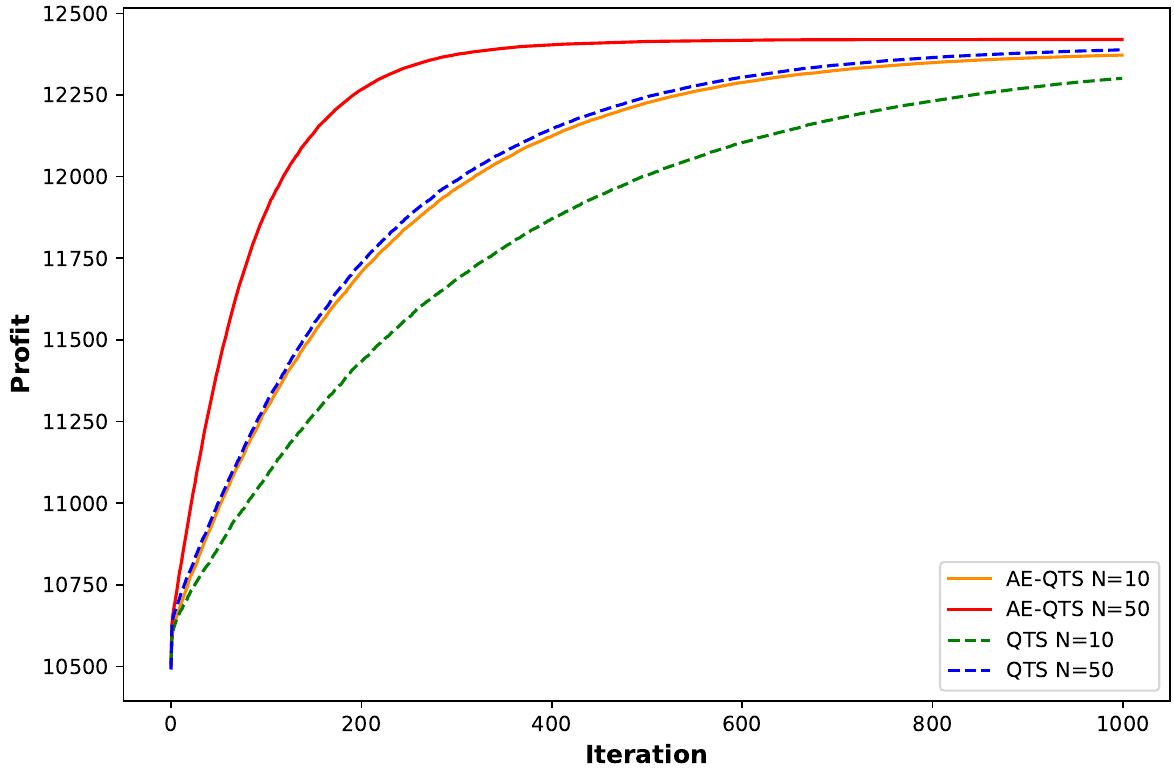}
            \caption{QTS vs AE-QTS for different population sizes ($N$).\newline\hspace*{0.5cm}The numbers of evaluation ($N*iteration$) are 10000 \newline\hspace*{0.5cm}and 50000, respectively.}
            \label{fig:QTS_AEQTS}
        \end{subfigure}
        \hfill
        \begin{subfigure}[b]{0.45\textwidth}
            \centering
            \includegraphics[width=\textwidth]{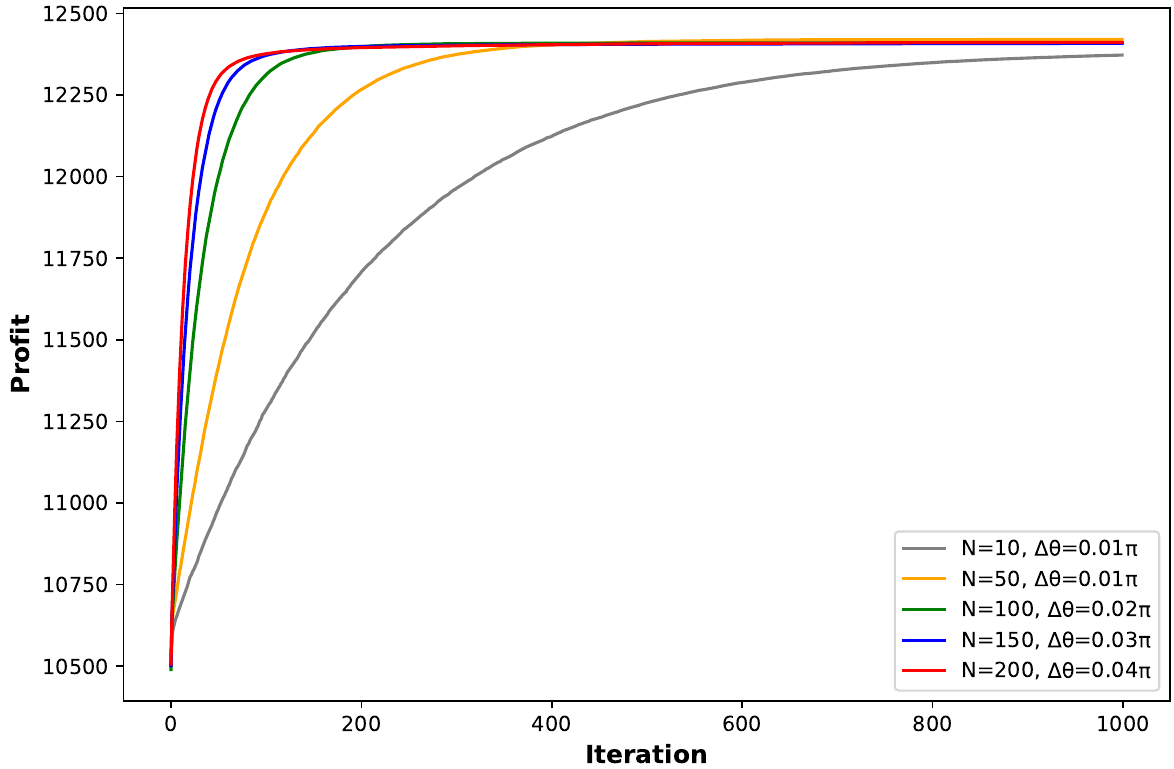}
            \caption{The relationship between population size ($N$) and $\Delta\theta$.\newline\hspace*{0.5cm}The numbers of evaluation ($N*iteration$) are 10000,\newline\hspace*{0.5cm}50000, 100000, 150000, and 200000, respectively.}
            \label{fig:n_theta}
        \end{subfigure}
        \caption{Case III with 2000 items.}
        \label{fig:diff_comparison}
    \end{figure*}

    Since all algorithms in all cases exhibit the same characteristics, we only conducted experiments on CASE III with 2000 items. We modified the $population\_size$ values in QTS \cite{chou2011quantum,chiang2014quantum} and AE-QTS, recording the convergence situations for population sizes of 10 and 50, as shown in Fig.~\ref{fig:QTS_AEQTS}. It was observed that AE-QTS with a population size of 10 performed almost as well as QTS \cite{chou2011quantum,chiang2014quantum} with a population size of 50. Furthermore, when the population size of AE-QTS was increased to 50, the performance enhancement was more significant compared to the increase from 10 to 50 in QTS \cite{chou2011quantum,chiang2014quantum}. This indicates that our strategy effectively incorporates all population information into the quantum state updates, resulting in a noticeable improvement in performance.

    Subsequently, we modified the population size and $\Delta\theta$ values, hoping to increase the convergence speed of the algorithm by increasing the $\Delta\theta$ value. We found that in experiments with larger population sizes, increasing $\Delta\theta$ did indeed increase the convergence speed, as shown in Fig.~\ref{fig:diff_comparison}. Moreover, the final convergence results were similar and did not fall into local optima prematurely. In contrast to QEA \cite{han2002quantum} and QTS, increasing $\Delta\theta$ will cause these algorithms to converge prematurely and fall into local optima. Therefore, the value of $\Delta\theta$ is almost problem-dependent and requires parameter tuning. In contrast, Fig.~\ref{fig:diff_comparison} shows that the impact of $\Delta\theta$ on AE-QTS appears relatively small.

    Lastly, we added a variable to all experiments to record the time of the last update of the global optimum. After averaging the final values, we obtained the final improvement performance. The main performance improvements are shown in Table~\ref{tbl:comparison}. Therefore, it can be deduced that the ``amplitude-ensemble'' mechanism can increase the performance of QTS \cite{chou2011quantum,chiang2014quantum} by approximately 34.74\%, 30.99\%, and 20.62\% for problems with 100, 250, and 500 items, respectively. Although the performance improvement may decrease with the complexity of the problem, the minimum value can still be maintained at around 20\%. Moreover, it can be seen from the figures of the experimental results that AE-QTS is clearly better than QTS \cite{chou2011quantum,chiang2014quantum} in the convergence process. In addition, the difference is more pronounced in complex problems, which also shows the advantages of this method.

    \begin{table}[ht]
        \centering
        \caption{The performance improvement comparison}
        \begin{tabular}{rllll}
        \hline
        \multicolumn{1}{|l|}{CASE \#}              & \multicolumn{1}{l|}{\begin{tabular}[c]{@{}l@{}}Num.\\ of item\end{tabular}} & \multicolumn{1}{l|}{QTS\textsuperscript{1}}    & \multicolumn{1}{l|}{AE-QTS\textsuperscript{1}} & \multicolumn{1}{l|}{PoI}     \\ \hline
        \multicolumn{1}{|r|}{\multirow{3}{*}{I}}   & \multicolumn{1}{l|}{100}                                                    & \multicolumn{1}{l|}{359.4} & \multicolumn{1}{l|}{219.55} & \multicolumn{1}{l|}{38.91\%} \\ \cline{2-5} 
        \multicolumn{1}{|r|}{}                     & \multicolumn{1}{l|}{250}                                                    & \multicolumn{1}{l|}{660.34} & \multicolumn{1}{l|}{441.53} & \multicolumn{1}{l|}{33.14\%} \\ \cline{2-5} 
        \multicolumn{1}{|r|}{}                     & \multicolumn{1}{l|}{500}                                                    & \multicolumn{1}{l|}{869.34} & \multicolumn{1}{l|}{703.77} & \multicolumn{1}{l|}{19.05\%} \\ \hline
        \multicolumn{1}{|r|}{\multirow{3}{*}{II}}  & \multicolumn{1}{l|}{100}                                                    & \multicolumn{1}{l|}{382.49} & \multicolumn{1}{l|}{259.58} & \multicolumn{1}{l|}{32.13\%} \\ \cline{2-5} 
        \multicolumn{1}{|r|}{}                     & \multicolumn{1}{l|}{250}                                                    & \multicolumn{1}{l|}{749.9} & \multicolumn{1}{l|}{530.09} & \multicolumn{1}{l|}{29.31\%} \\ \cline{2-5} 
        \multicolumn{1}{|r|}{}                     & \multicolumn{1}{l|}{500}                                                    & \multicolumn{1}{l|}{958.3} & \multicolumn{1}{l|}{821.56}    & \multicolumn{1}{l|}{14.27\%} \\ \hline
        \multicolumn{1}{|r|}{\multirow{3}{*}{III}} & \multicolumn{1}{l|}{100}                                                    & \multicolumn{1}{l|}{207.89} & \multicolumn{1}{l|}{138.95} & \multicolumn{1}{l|}{33.16\%} \\ \cline{2-5} 
        \multicolumn{1}{|r|}{}                     & \multicolumn{1}{l|}{250}                                                    & \multicolumn{1}{l|}{435.25}    & \multicolumn{1}{l|}{302.36} & \multicolumn{1}{l|}{30.53\%} \\ \cline{2-5} 
        \multicolumn{1}{|r|}{}                     & \multicolumn{1}{l|}{500}                                                    & \multicolumn{1}{l|}{711.97} & \multicolumn{1}{l|}{508.79} & \multicolumn{1}{l|}{28.54\%} \\ \hline
        \multicolumn{5}{|l|}{\begin{tabular}[c]{@{}l@{}}The average improvement ratios: \\ 34.74\% (100 items), 30.99\% (250 items), 20.62\% (500 items) \\ The total average improvement is 28.78\%\end{tabular}}                                                        \\ \hline
        \multicolumn{5}{l}{Note\textsuperscript{1}: the average iteration for the last update of the global optimal.} \\
        \multicolumn{5}{l}{\textbf{Note: the number of evaluations all are 10,000.}} \\
        \multicolumn{5}{l}{PoI: percentage of improvement}
                                                                                 
        \end{tabular}
        \label{tbl:comparison}
    \end{table}

\section{Conclusion}
    In this study, we successfully incorporated the population concept of the Grover Search Algorithm \cite{grover1996fast} into all the qubits of AE-QTS, achieving at least approximately an average 20\% increase compared with QTS \cite{chou2011quantum,chiang2014quantum}. In complex problems, the quality of the solutions found by AE-QTS also manages to exceed that of QTS \cite{chou2011quantum,chiang2014quantum}. Given that QTS's inherent performance already surpasses typical metaheuristic methods, this increment in efficiency is indeed a remarkable contribution.

    In addition, the improvement method proposed in this study does not increase the complexity of QTS itself; only simply modifies the core of QTS \cite{chou2011quantum,chiang2014quantum}. Therefore, it can be achieved by simply modifying the methods that have been written using QTS \cite{chou2011quantum,chiang2014quantum}. This means that all methods that adopt QTS \cite{chou2011quantum,chiang2014quantum} as the core can obtain this performance improvement.

    In conclusion, our study maintains the simplicity of the QTS method \cite{chou2011quantum,chiang2014quantum}, adding no additional parameters. The new quantum state update method is devoid of ambiguous spaces, avoiding issues of ambiguity in programming. This makes it exceptionally easy to implement and adaptable to various problems.

\section*{Acknowledgment}
Thanks to the efforts of all team members, authorship order does not truly reflect the extent of their contributions. This work was supported by the National Science and Technology Council, Taiwan, R.O.C., under Grants 111-2222-E-197-001-MY2.

\bibliographystyle{IEEEtran.bst}
\bibliography{main}
\balance

\end{document}